\documentclass{PoS}

\title{Episodic Torque-Luminosity Correlations and Anticorrelations
of GX 1+4}

\ShortTitle{Episodic Torque-Luminosity Correlations and Anticorrelations
of GX 1+4}

\author{\speaker{M. M. Serim}\\
        Middle East Technical University\\
        E-mail: \email{muhammed@astroa.physics.metu.edu.tr}}
        
\author{\c{S}. \c{S}ahiner\\
       Middle East Technical University \\
        E-mail: \email{seyda@astroa.physics.metu.edu.tr}}

\author{D. \c{C}erri-Serim\\
        Middle East Technical University\\
        E-mail: \email{danjela@astroa.physics.metu.edu.tr}}
\author{S. \c{C}. \.{I}nam\\
        Ba\c{s}kent University\\
        E-mail: \email{inam@baskent.edu.tr}}
\author{A. Baykal\\
        Middle East Technical University\\
        E-mail: \email{altan@astroa.physics.metu.edu.tr}}

\abstract{We analyse archival CGRO-BATSE X-ray flux and spin frequency measurements of GX 1+4 over a time span of 3000 days. We systematically search for time dependent variations of torque luminosity correlation. Our preliminary results indicate that the correlation shifts from being positive to negative on time scales of few 100 days.  }

\FullConference{11th INTEGRAL Conference Gamma-Ray Astrophysics in Multi-Wavelength Perspective,\\
		10-14 October 2016\\
		Amsterdam, The Netherlands}

\begin{document}

\section{Introduction}
GX 1+4 is an accretion powered pulsar of which optical companion is M6III type M giant V2116 Oph with slow stellar wind (Chakrabarty \& Roche 1997; Davidsen et al. 1977;
Hinkle et al. 2006). It was discovered with approximately $\simeq$138 s pulsations (Lewin et al. 1971)
and found to be spinning-up till it became undetectable during 1980s (Doty et al. 1981).
Then the source detected again during late 1980s and it is found to be spinning-down indicating that source underwent a torque reversal during undetectable era (White et al. 1983;Makishima et al. 1988).
The source shows an overall spin-down trend ever since (Gonzales-Galan et al. 2012).

Previous studies of GX 1+4 revealed correlations between frequency derivative and X-ray luminosity (Paul et al. 1997; Chakrabarty et al. 1997). Chakrabarty et al. (1997) investigated the torque and luminosity behaviour for 2000 days of BATSE measurements and reported that the frequency derivative and X-ray luminosity episodically shows both positive and negative correlations.
\section{Observations}
GX 1+4 was continuously monitored with Burst and Transient Source Experiment (BATSE) between MJD 48373 and 51348 for  a time span of approximately 3000 days.
We retrieved pulse frequency, frequency derivative and X-ray pulsed flux measurements of the observations from public archive of BATSE Pulsar Team\footnote{https://gammaray.nsstc.nasa.gov/batse/pulsar/}.  The pulse frequency measurements are initially carried out with searching the pulse signal via fast fourier transform. When the pulse frequency of a source is already known, more accurate measurements are implemented via other timing methods (such as epoch folding and pulse phase). The measurements are made with 5 days sampling and flagged either Y or N depending on the whether confidence level of the measurement is above the noise level (see BATSE pulsar team website for more information). In this study, we only used measurements flagged Y.
The retrieved data are investigated for frequency derivative and X-ray pulsed flux correlations. In this work, we seek for such correlations on shorter time scales.
\section{Time Dependent Torque-Luminosity Correlation}
We analyse the data systematically for time dependent variations of correlation and anti-correlation episodes. Hence, we blindly search  for  correlation by calculating Pearson correlation coefficient between two parameters for time row number 0 to N. When the correlation coefficient systematically changes after row n, we accept the Pearson coefficient of this interval to be pre-break down value (between 0 and n) and repeat same procedure each time for next interval between row n and (n+N). The correlation coefficient value breaks down on time scales of 100 days (Serim et al 2017). Furthermore, correlations are generally weak (Pearson coefficient is ranging between -0.6 and 0.6). In figure 1, we represent frequency derivative and X-ray flux of each corresponding interval. 
\begin{figure}[ht]
\centering
\includegraphics[width=1.0\linewidth]{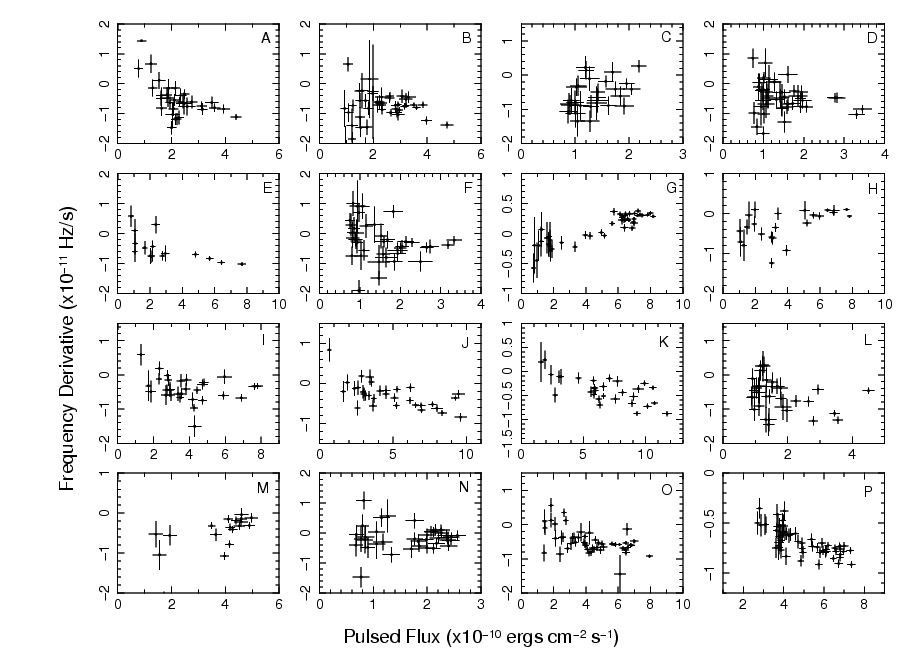} 
\caption{Frequency derivative vs pulsed flux correlation of each interval. MJD ranges of intervals are (A) 48364-48508; (B) 48512-48724; (C) 48728-48944; (D) 48948-49176; (E) 49180-49292; (F) 49296-49580; (G) 49584-49764; (H) 49768-49872; (I) 49876-49992; (J) 49996-50124; (K) 50128-50244; (L) 50276-50628; (M) 50632-50708; (N) 50720-50920; (O) 50924-51128; (P) 51132-51348, respectively. }
\end{figure}

\section{Conclusion}
We investigate daily CCRO-BATSE pulsed X-ray flux and pulse frequency derivative measurements of GX 1+4 in 20-60 keV energy band and search for time resolved correlations between these parameters. This correlation can be accepted as a sign of torque-luminosity variations under the assumption that bolometric luminosity is correlated with X-ray flux. We found that the source seems to undergo episodic variations of the correlation on time scales of $\simeq 100-200$ days.

An anti-correlation between torque and luminosity is hard to explain by standard prograde accretion disk models. However, formation of a transient prograde/retrograde disk by a stellar wind may account for both variation of correlation and observed timing noise behaviour (GX 1+4:Bildsten et al. 1997; Nelson et al. 1997; Serim et al. 2017, OAO 1657-415: Baykal 1997; Baykal 2000). The episodic interchange of correlations might be an indication of prograde/retrograde disk formation on time scales of few 100 days.

We acknowledge support from T\"{U}B\.{I}TAK, the Scientific and Technological Research Council of Turkey through the research project MFAG 114F345.

\end{document}